\documentclass{article}

\PassOptionsToPackage{numbers, sort, compress}{natbib}
\usepackage[final]{neurips_2022_ml4ps}
\usepackage[utf8]{inputenc} % allow utf-8 input
\usepackage[T1]{fontenc}    % use 8-bit T1 fonts
\usepackage{hyperref}       % hyperlinks
\usepackage{url}            % simple URL typesetting
\usepackage{booktabs}       % professional-quality tables
\usepackage{amsfonts}       % blackboard math symbols
\usepackage{nicefrac}       % compact symbols for 1/2, etc.
\usepackage{microtype}      % microtypography
\usepackage{xcolor}         % colors

\usepackage{mathtools}
\usepackage{xspace}
\usepackage{amsmath}
\usepackage{bbm}
\usepackage{amssymb}
\usepackage{pifont}
\usepackage{caption}
\usepackage{subcaption}
\usepackage{graphicx}
\usepackage{wrapfig}
\usepackage{enumitem}
\usepackage{multirow}
\usepackage{comment}
\usepackage[acronym]{glossaries}

\usepackage[style=american]{csquotes}\MakeOuterQuote{"}
\usepackage{cleveref}
\usepackage{bm}
\usepackage{dsfont}
\usepackage{physics}
\usepackage{siunitx}
\usepackage{interval}

\usepackage{fancyvrb}
\NewDocumentCommand{\defineverb}{m m}{
	\newcommand{#1}{\UseVerb{\string#1}\xspace}
	\DefineShortVerb{\|}\SaveVerb{\string#1}|#2|\UndefineShortVerb{\|}
}
\defineverb{\pytorch}{PyTorch}

\newcommand{\eg}{e.g.\@\xspace}
\newcommand{\ie}{i.e.\@\xspace}

\newcommand{\iid}{i.i.d.\@\xspace}

\newcommand{\AstroDDPM}{AstroDDPM\xspace}
\newcommand{\PROBES}{PROBES\xspace}
\newcommand{\SLACS}{SLACS\xspace}
\newcommand{\BELLS}{BELLS\xspace}

\newcommand{\reals}{\mathbb{R}}
\newcommand{\Identity}{\mathds{1}}
\newcommand{\Normal}{\mathcal{N}}
\newcommand{\given}{\,\vert\,}
\newcommand{\transpose}{\top}

\newcommand{\n}{n}
\newcommand{\m}{m}

\newcommand{\x}{\mathbf{x}}
\newcommand{\y}{\mathbf{y}}
\renewcommand{\H}{\mathbf{H}}
\newcommand{\z}{\mathbf{z}}

\newcommand{\lensingobs}{\y_{\text{obs}}}
\newcommand{\lensingsrc}{\x_{\text{src}}}
\newcommand{\lensingH}{\H_{\text{lens}}}
\newcommand{\lensingnoise}{\z}

\newcommand{\nnparams}{\Theta}
\newcommand{\meanfunc}{f_{\nnparams}}
\newcommand{\q}{\operatorname{q}}
\newcommand{\p}{\operatorname{p}}

\newcommand{\noiseq}[1][t]{\operatorname{q\ifblank{#1}{}{^{(#1)}}}}
\newcommand{\denoisep}[1][t]{\operatorname{p\ifblank{#1}{}{^{(#1)}}_{\nnparams}}}

\newcommand{\noise}{\sigma_{y}}

\newcommand{\U}{\mathbf{U}}
\renewcommand{\S}{\mathbf{S}}
\renewcommand{\si}{s_i}
\newcommand{\V}{\mathbf{V}}
\newcommand{\Vh}{\mathbf{V}^{\transpose}}

\newcommand{\roti}[1]{\bar{#1}^{(i)}}
\newcommand{\xbar}{\bar{\x}}
\newcommand{\xbari}{\roti{\x}}
\newcommand{\xrec}{\xbari_{\nnparams, t}}

\newcommand{\ybari}{\roti{\y}}

\newacronym{nn}{NN}{neural network}
\newacronym{elbo}{ELBO}{evidence lower bound}
\newacronym{ddpm}{DDPM}{denoising diffusion probabilistic model}
\newacronym{ddim}{DDIM}{denoising diffusion implicit model}
\newacronym{ddrm}{DDRM}{denoising diffusion restoration model}
\newacronym{svd}{SVD}{singular value decomposition}

\title{Strong-Lensing Source Reconstruction with Denoising Diffusion Restoration Models}
\author{
    Konstantin Karchev\\
    Theoretical and Scientific Data Science\\
    SISSA, Trieste, Italy\\
    \texttt{kkarchev@sissa.it}
    \And
    Noemi Anau Montel\\
    GRAPPA Institute\\ 
    University of Amsterdam, The Netherlands\\ 
    \texttt{n.anaumontel@uva.nl}
    \And
    Adam Coogan\\
    Ciela -- Computation and Astrophysical Data Analysis Institute, Montréal, Quebec, Canada\\
    Département de Physique, Université de Montréal, Canada\\
    Mila -- Quebec AI Institute, Montreal, Canada\\
    \texttt{adam.coogan@umontreal.ca}
    \And
    Christoph Weniger\\
    GRAPPA Institute\\ 
    University of Amsterdam, The Netherlands\\
    \texttt{c.weniger@uva.nl}
}

\begin{document}
\maketitle

\begin{abstract}
%   Strong gravitational lensing on galactic scales has the potential to be the scene for the discovery of the nature of dark matter through characterisation of the abundance of substructure.
    Analysis of galaxy--galaxy strong lensing systems is strongly dependent on any prior assumptions made about the appearance of the source. Here we present a method of imposing a data-driven prior / regularisation for source galaxies based on \acrfullpl*{ddpm}. We use a pre-trained model for galaxy images, \AstroDDPM, and a chain of conditional reconstruction steps called \acrfull*{ddrm} to obtain samples consistent both with the noisy observation and with the distribution of training data for \AstroDDPM. We show that these samples have the qualitative properties associated with the posterior for the source model: in a low-to-medium noise scenario they closely resemble the observation, while reconstructions from uncertain data show greater variability, consistent with the distribution encoded in the generative model used as prior.
\end{abstract}

\section{Introduction}
\label{sec:intro}

Gravitational lensing, the phenomenon of light bending trajectory under the influence of gravitating mass, has enabled progress in diverse areas of physics: from discovering some of the furthest observed galaxies in the Universe \citep{egsy8p7,Naidu_2022} and analysing them \citep[\eg][]{GLASS} to inferring the dark matter content of clusters and its distribution on galactic and sub-galactic scales \citep{Dalal_2002,Vegetti_2014,Gilman_2020,Hsueh_2020,Meneghetti_2020,Nightingale_2022}, including detections of individual light dark matter halos without luminous counterparts \citep{Vegetti_2010b,Vegetti_2012,Hezaveh_2016}, and measuring the Hubble constant \citep{suyu2020holismokes,birrer2020tdcosmo}. Critical in most endeavours is the ability to model the complex morphology of lensed sources, either as a goal in and of itself, or in order to disentangle their surface brightness inhomogeneity from perturbations in the lens.

Existing strong-lensing source models can be roughly classified in four categories with increasing complexity: analytic parametrisations like the S\'ersic profile \citep{Sersic_1963,brownstein2011boss}; regularised pixellation of the source plane \citep{Warren_Dye_2003,Suyu_2006,Vegetti_Koopmans_2009,Nightingale_Dye_2015,Nightingale_Dye_Massey_2018} (where the regularisation can be implicit in the use of \eg a Gaussian process prior \citep{Karchev_Coogan_Weniger_2022} or continuous neural fields \citep{Mishra-Sharma_Yang_2022}); basis function regression onto \eg wavelets \citep{Galan_2021} or shapelets \citep{Birrer2015-xb,Birrer2018-rq}; and deep learning approaches with \eg recurrent inference machines \citep{Morningstar_2018,Morningstar_2019,Adam22RIM} or variational autoencoders \citep{Chianese_2020}. While the former three categories are based on specific model assumptions, the deep-learning approaches are data-driven: it aims to learn from observations what typical galaxies look like and steer reconstructions appropriately. And while the current set of galaxy--galaxy strong lensing observations number on the order of hundreds, mainly coming from dedicated campaign like \SLACS \citep{SLACS-1,SLACS-5,SLACS-13} and \BELLS \citep{BELLS-1,BELLS-3}, future general-purpose cosmological surveys are expected to deliver \emph{hundreds of thousands} more \citep{Collett_2015}, which underlines the need for fast and robust inference methodologies.

In this work we demonstrate galaxy--galaxy strong-lensing source reconstruction using denoising diffusion, the state-of-the-art deep-learning generative technique at the core of recent striking text-to-image models like DALL-E~2 \citep{dalle2}. The aim of any generative model (see \eg \citet{generative-review} for a "recent" review) is to learn from (usually very high-dimensional) data an approximation to the underlying distribution from which it has been drawn and enable easy sampling of new high-fidelity examples. \Glspl*{ddpm}, introduced by \citet{Sohl-Dickstein_2015} and elaborated by \citet{ddpm}, achieve this by learning to reverse the gradual degradation of an input with random noise. By carefully designing both the noising and denoising processes, one can arrive at a particularly simple structure of the overall model, where a \gls*{nn} is trained to predict the mean of a Gaussian used in denoising.

We use a \gls*{ddpm} pre-trained on galaxy images called \AstroDDPM \citep{astroddpm} and a modified sampling procedure called \gls*{ddrm} \citep{ddrm} to condition the generation on a particular strong-lensing observation. We verify that this results in samples that exhibit desirable properties of the Bayesian posterior: when the noise in the observation is low, reconstructions follow it closely, while when noise is significant, samples are dictated by the data-driven prior encoded in \AstroDDPM and show significant variation while still being consistent with the data. We expect the denoising diffusion approach to source reconstruction to prove instrumental in generating constrained training examples for simulation-based inference of dark matter substructure properties.

% Precision analysis of strong gravitational lensing images has proven a powerful tool to constrain dark matter models \citep{} and detect dark matter substructures \citep{Vegetti_2010a,Vegetti_2010b, Vegetti_2012, Hezaveh_2016a}.
% A critical point for these analyses is the  

% Talk about existing methods and their drawbacks? low-dimensional parametric models (e.g. Sérsic), pixellation of the source plane, source modeling with basis functions (e.g. shapelets), Gaussian processes.

% Recent years have seen the enormous success of generative models in learning any kind of data distribution using unsupervised learning. In a nutshell, generative models can be trained on a large collection of content (\eg images, text, sounds) in order to approximate the true data distribution and then applied to generate new contents that resemble the source data with some variations.

% In this work, we focus on diffusion models, a particular class of generative models, for strong-lensing source galaxy reconstruction. First proposed by \citet{Sohl-Dickstein_2015}, \gls*{ddpm} are latent-variable models that have proven capable to synthesize high-quality images sampling from a learnt distribution by reversing a gradual noising process. Specifically, we propose to use \gls*{ddrm} to generate high-quality source-galaxy samples from a model conditioned on the lensing observation. 

% Talk about the importance of targeted data and being able to have samples that are faithful to the observation?

\section{Background}
\label{sec:diffusion}

\subsection{Galaxy--galaxy strong gravitational lensing}\label{sec:lensing}

Galaxy--galaxy strong gravitational lenses are usually modelled in the \emph{thin-lens approximation} whereby all observed light is assumed to have originated from a specified \emph{source plane} and been deflected by mass concentrated in a \emph{lens} (or \emph{image}) \emph{plane} located between the source and the observer. Thin-lensing is entirely defined by the field of \emph{deflection angles}, which is calculated from, and thus encodes, the mass distribution in the lens plane: see \citet{Meneghetti-lensing} for full details.

% Gravitational lensing is the effect of bending the path of light due to the gravitational influence of a nearby mass distribution. The setting that we consider (galaxy--galaxy strong lensing) is usually modelled with the \emph{thin-lens approximation}, whereby all the light is assumed to be coming from a specified \emph{source plane}, and all the lensing mass is concentrated in a \emph{lens (or image) plane} located between the source and the observer.
% The angular coordinate in the source plane from which a ray of light originated, $\coordsrc$, is then related to the direction from which it \emph{appears} to be coming, $\coordimg$, and the displacement field, $\disp(\coordimg)$, via the \emph{ray-tracing equation}: $\coordsrc(\coordimg) = \coordimg - \disp(\coordimg)$, where all angles are assumed to be small. The displacement field $\disp(\coordimg)$, in turn, is determined from the distribution of mass over the whole lens plane: we refer the reader to \citet{Meneghetti-lensing} for full details on the physics of gravitational lensing.

Importantly, gravitational lensing preserves surface brightness since it does not create or destroy photons, and so the observed flux in the image plane is simply the flux of the source at the origin of the ray.
% : $\flux(\coordimg) = \flux(\coordsrc(\coordimg))$.
This means that lensing is a linear process, and source reconstruction can be phrased as a linear inversion problem, if the source is modelled on a (possibly irregular) grid, as recognised by \citep{Warren_Dye_2003,Suyu_2006}. Since the grid can be made as fine as possible, while the observations have a fixed (usually coarse) resolution, and due to the almost complete degeneracy between lens and source, the \emph{regularisation} and/or \emph{Bayesian prior} on the source model has a crucial role both for the quality of the reconstruction, and for subsequent analysis performed with it (\eg lens substructure inference).
% In the language of Bayesian inference, the regularisation of the source appears as a prior for what sources look like: \eg smooth or inhomogeneous on certain scales, with or without spiral structure, \etc.

Usually, the lensing configuration is a priori unknown (or weakly constrained by observations of the light of the lens galaxy) and often itself a target for inference. In this work, however, we focus only on source reconstruction, so we assume we know the details of the lens perfectly: \ie we know both the mass configuration and the light of the lensing galaxy, which we can perfectly subtract from the observation. Our model, then, can be stated as
\begin{equation}\label{eqn:linear-lensing}
    \lensingobs = \lensingH \lensingsrc + \lensingnoise,
\end{equation}
where $\lensingobs \in \reals^\n$ is the observed image (flattened to a vector), $\lensingsrc \in \reals^\m$ is the gridded source model, and $\lensingnoise$ is observational noise, assumed \iid Gaussian in each pixel: $\lensingnoise \sim \Normal(\mathbf{0}, \noise^2 \Identity)$. The $\n$-by-$\m$ matrix $\lensingH$ encodes the lensing distortions, instrumental effects (such as a point-spread function), and interpolation of $\lensingsrc$ across the grid on which it is defined. We use a ray-tracing code built with \pytorch in order to calculate $\lensingH$ with automatic differentiation of a forward simulation (since \cref{eqn:linear-lensing} is linear, the particular values used for $\lensingsrc$ in the forward pass are immaterial).

\subsection{Denoising diffusion probabilistic models (DDPM)}

\Glspl*{ddpm} are a class of unsupervised density estimation techniques that aim to learn the underlying distribution $\q(\x)$ of data $\{\mathbf{X}_i\}_{i=1}^{N}$ in a way that is then easy to sample from. They achieve this by introducing $T$ \emph{latent spaces} $\x_t$ for $t = 1:T$, which are modelled in two ways: via a forward
(diffusion) and a reverse (generative) processes. The forward process is a Markov chain that slowly adds Gaussian noise with increasing variance\footnote{
    There are two conventions for scheduling the noise, termed "variance exploding" and "variance preserving". Here, we present the latter, which is used to train \AstroDDPM, even though the \gls*{ddrm} implementation we use is variance exploding: for the conversion between the two, see Appendix B of \citet{ddrm}.
} $\beta_t$ to the initial data point: $\noiseq(\x_{t} \given \x_{t-1}) = \Normal(\sqrt{1-\beta_t}\x_{t-1}, \beta_t \Identity)$, ending up with essentially pure noise. The model then learns the inverse (iteration-dependent) \emph{denoising} operation $\denoisep(\x_{t} \given \x_{t+1})$, which is usually again modelled as a Gaussian with pre-determined variance, whose mean is provided by a neural network $\meanfunc(\x_{t+1})$. Optimisation is performed with gradient ascent on the \gls*{elbo} of this model: a measure of the similarity over the training data between the forward and reverse distributions of the latent spaces:
\begin{equation}
    \noiseq[](\x_{0:T}) = \noiseq[](\x_0) \prod_{t=1}^{T}\noiseq(\x_t\given\x_{t-1})
    \qquad
    \leftrightarrow
    \qquad
    \denoisep[](\x_{0:T}) = \denoisep[T](\x_T) \prod_{t=0}^{T-1} \denoisep(\x_t\given\x_{t+1}).
\end{equation}
While $\noiseq[](\x_0)$ is approximated with the training data, $\denoisep[T](\x_T)$ is set to a Gaussian with unit variance,
% \kk{$\beta_T$ - CW: Actually one}
so that one can draw pure noise and iteratively denoise it to obtain a new sample for $\x_0$.

\subsection{\texorpdfstring{\acrshort*{ddpm}}{DDPM} as a prior: \texorpdfstring{\acrfull*{ddrm}}{DDRM}}
We would like to use the learnt approximation to $\q(\x)$ as a prior for the linear inversion problem stated in \cref{eqn:linear-lensing}, \ie sample from the posterior $\p(\x\given\y) \propto \q(\x) \p(\y\given\x)$ with a Gaussian likelihood $\p(\y\given\x) = \Normal(\y \given \H\x, \noise^2\Identity)$. This can be achieved by conditioning $\noiseq$ and $\denoisep$ on $\y$ and training bespoke density estimators (\ie denoisers) for each observation. However, this is obviously very computationally expensive and does not scale to analyses of multiple systems that differ only by the lensing matrix $\lensingH$ but not by the assumed prior on $\x$.

\citet{snips,ddrm} showed that, under certain conditions, a \emph{pre-trained} \gls*{ddpm} model can be indeed used as a prior in a linear inversion model in order to sample from a \emph{constrained} generative process $\denoisep[](\x_{0:T}\given\y) = \denoisep[T](\x_T\given\y) \prod_{t=0}^{T-1} \denoisep(\x_t\given\x_{t+1}, \y)$. Their solution is expressed in the singular space of $\H$ and therefore starts with computing its \gls*{svd}:
\begin{equation}
    \H = \U \S \Vh,
\end{equation}
and applying the transformations $\bar{\y} \equiv \S^{+}\U^{\transpose}\y$ (where $\S^{+}$ is a Moore--Penrose pseudo-inverse) and $\bar{\x} \equiv \Vh \x$. The sampling procedure then considers separately components which are constrained by the data---\ie those that have a positive singular value $\si > 0$---from those that are not ($\si = 0$). Crucially, in the initial iterations, in which the noise level in the Markov chain is larger than the observational uncertainty, the denoising procedure is steered towards the observation with a weight $\eta_b = 2\sigma_t^2 / (\sigma_t^2 + \noise^2 / \si^2)$, where $\sigma_t^2$ is the accumulated noise variance at step $t$. At each step, the pre-trained denoiser $\meanfunc$ is only used to calculate the mean for the following step: $\x_{\nnparams, t} = \meanfunc(\x_{t+1})$, which is then rotated into $\xbar_{\nnparams, t} \equiv \Vh \x_{\nnparams, t}$. \Gls*{ddrm} has one hyperparameter, $\eta$, which relates to the specific way the denoising network has been trained and also influences the amount by which denoising is steered towards the observation. The specific form of the \acrshort*{ddrm} updates is given in \cref{eqn:ddrm-1,eqn:ddrm-2} in \cref{apx:ddrm}.

\section{Demonstration on mock data and discussion}\label{sec:results}

In this \namecref{sec:results}, we apply \gls*{ddrm} to realistic mock observations of galaxy--galaxy strong lensing. We use the \AstroDDPM\footnote{
    \url{https://github.com/Smith42/astroddpm}, released under the AGPL-3.0 open-source license
} network \citep{astroddpm}, pre-trained on the \PROBES dataset \citep{Stone_2019,Stone_2021}, which contains \num{1962} images of late-type galaxies that exhibit fine structure and details. We note that the \PROBES dataset may not be representative of high-redshift source galaxies appearing in strong lenses, and so future analyses should check for possible biases due to the choice of training data.
% \Citet{astroddpm} show that their model generalises well and produces samples that are not simple reproductions of the training images, and so we use it both to generate the true source for the mock data, and for inference, extracting the denoiser $\meanfunc$ from the publicly available code.
% \AstroDDPM is
% It has been trained on multi-channel (colour) images, and the encoded prior has a strong preference for red galaxies.
Since \AstroDDPM is a multi-channel model (with channels corresponding to the $g$, $r$ and $z$ photometric bands), in order to avoid biases, we simulate and analyse multi-channel images as well, considering an independent likelihood as in \cref{eqn:linear-lensing} for each channel. We note, still, that it is possible to include any linear operation on the channels (\eg "selecting a channel" or "averaging channels") inside $\lensingH$.

We set up a uniform source-plane grid spanning $\SI{1 x 1}{\arcsecond}$ with $\m = \num{256 x 256}$ pixels, to match \AstroDDPM. We choose a resolution of \SI{0.05}{\arcsecond} and $\n = \num{50 x 50}$ pixels in the image plane, as appropriate for Hubble observations
% compiled in \SLACS.
and set up a lens configuration so that multiple images are formed.
% As a lensing configuration we take a standard singular power-law ellipsoid (SPLE) smooth main lens with additional external shear and set parameters inspired by the \SLACS lenses, such that multiple images of the source are formed.
In the end, only \num{\sim 1000} image pixels trace back to the source grid.
% , so this should be considered the size of our data (in lensing analyses the dark parts of the image are usually masked out).
As a source image we use an unconditioned sample from \AstroDDPM.

To account for integrating the flux within an pixel, which is especially important in highly-magnified parts of the image, when calculating $\lensingH$, we simulate at a \num{10}-times higher image resolution and then downsample with local averaging. As a cross-check, we have verified that multiplying $\lensingH \lensingsrc$ matches the full simulation output to numerical precision. Finally, we add independent pixel noise. We test two settings: a "medium-noise" regime where $\noise$ is set to $1/30$ of the brightest image pixel, consistent with typical \SLACS lenses; and a "high-noise" regime with peak signal-to-noise ratio of only \num{6} in order to verify that the \gls*{ddrm} produces enough variation when the observation is not very constraining.

\begin{figure}[t]
    \centering
    \includegraphics[width=\linewidth]{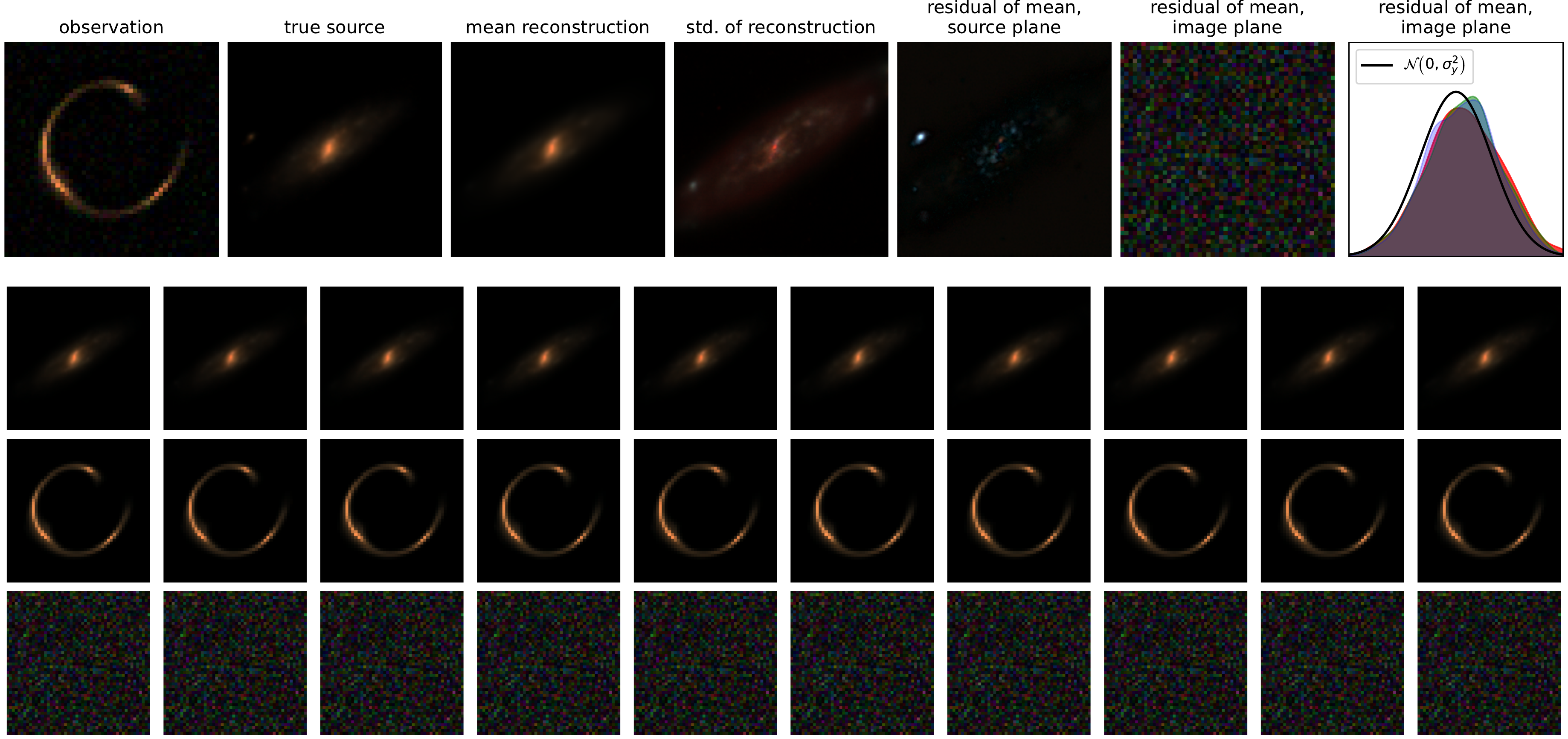}
    \caption{Top: from left to right, the mock observation, $\y$ (with a medium noise level), the true source, $\x$ (an unconstrained sample from \AstroDDPM), the mean and standard deviation of \num{100} posterior samples from \gls*{ddrm}, $\x_{0, i} \sim \denoisep[](\x_0 \given \y)$, and the residual of the mean with respect to the true source and with respect to the observation in the image plane; finally, a histogram of the latter compared to a Gaussian. Bottom: each column is a random posterior sample (top row), which is then lensed to produce the respective noiseless image $\H\x_{0, i}$ (middle row). Shown (bottom row) are also the residuals between $\H\x_{0, i}$ and the observation. In residual plots, negative values in one channel are shown as positive values in the other two (red $\leftrightarrow$ cyan, green $\leftrightarrow$ magenta, blue $\leftrightarrow$ yellow), considering complementary colors as "negative".\label{fig:results-med}}
\end{figure}

Our main results are displayed in \cref{fig:results-med}. We set $\eta = \num{1}$ in accordance with the theorem of \citet{ddrm} and sample \num{100} realisations, which takes \SI{\sim 10}{\minute} on an NVIDIA A-100 GPU. We verify that in this medium-noise setting, the true source is reconstructed with high fidelity even from only \num{\sim 1000} pixels, owing to the multiple observed projections and properly taking integration within a pixel into account. Standardised residuals between the lensed mean reconstruction and the observation follow a unit normal distribution and show no structure or signs of bias. Individual samples vary to a degree appropriate for the observational noise.

If the noise level is increased by a factor of \num{5}, the reconstructions show accordingly higher variability (see \cref{fig:results-high} in the \namecref{apx:ddrm}). Conditioned samples now follow more closely the prior and display a larger variety of morphologies, sizes and brightnesses, while still being consistent with the observation (the residuals of the mean are still approximately normal), although reconstructions seem to be slightly dimmer in the red (brightest) channel.

\section{Conclusion}

We have shown that one can use a pre-trained denoising diffusion model and the procedure in \gls*{ddrm} to reconstruct source galaxies from noisy strong gravitational lensing data with high fidelity. The reconstructions exhibit a qualitative variability necessary for them to be interpreted as samples from a posterior for the source's appearance, and we intend to perform quantitative tests, \eg using the classification 2-sample test \citep{Friedman:2003id,lopez2017revisiting}, over a large number of mock observations to verify this. Such tests will also aid in setting the hyperparameter $\eta$ of \gls*{ddrm}. In future work we will also unite \gls*{ddrm} source reconstruction with a scheme for inferring the mass distribution of the lens galaxy, which defines the distortion matrix $\lensingH$. Finally, our intended application for the method presented here is for generating training data for simulation-based inference of dark matter substructure, which will require an extension of the methodology to handle the correlated and spatially varying noise present in real lensing observations. We are confident that even in its present form, strong-lensing source reconstruction with \glspl*{ddrm} can be a useful tool for astrophysics and cosmology.

\paragraph{Broader Impact}
\label{par:impact}
This work is focused on the precision analysis of strong gravitational lensing data via diffusion models, a class of generative models. 
Unfortunately, there are numerous well-known malicious uses of generative models (\eg sample generation techniques can be employed to produce fake images and videos that can impact people's lives). 
On the other hand, through our analysis of source reconstruction in strong lensing, we have proven diffusion models to be useful for solving high-dimensional Bayesian inference problems thanks to their ability to capture the statistics of natural datasets. 
Although we do not anticipate potential for misuse of the presented application, the usual caution has to be exercised when drawing scientific conclusions based on a complex analysis machinery.

%%%%%%%%%%%%%%%%%%%%%%%%%%%%%%%%%%%%%%%%%%%%%%%%%

\section*{Acknowledgments and Disclosure of Funding}

This work is part of a project that has received funding from the European Research Council (ERC) under the European Union’s Horizon 2020 research and innovation program (Grant agreement No.\ 864035 -- Undark). A.~C. acknowledges funding from the Schmidt Futures Foundation. K.~K. acknowledges the hospitality of the Galileo Galilei Institute, Florence.

% We acknowledge the use of the \texttt{python} \citep{python} modules, \texttt{matplotlib} \citep{matplotlib}, \texttt{numpy} \citep{numpy},  \texttt{scipy} \citep{scipy}, \texttt{PyTorch} \citep{pytorch}, \texttt{tqdm} \citep{tqdm}, and \texttt{jupyter} \citep{jupyter}.

%%%%%%%%%%%%%%%%%%%%%%%%%%%%%%%%%%%%%%%%%%%%%%%%%

\begingroup
\footnotesize
\setlength{\bibsep}{4pt}
\bibliographystyle{unsrtnat}
\bibliography{bibliography}

\begin{thebibliography}{48}
\providecommand{\natexlab}[1]{#1}
\providecommand{\url}[1]{\texttt{#1}}
\expandafter\ifx\csname urlstyle\endcsname\relax
  \providecommand{\doi}[1]{doi: #1}\else
  \providecommand{\doi}{doi: \begingroup \urlstyle{rm}\Url}\fi

\bibitem[{Zitrin} et~al.(2015){Zitrin}, {Labb{\'e}}, {Belli}, {Bouwens},
  {Ellis}, {Roberts-Borsani}, {Stark}, {Oesch}, and {Smit}]{egsy8p7}
Adi {Zitrin}, Ivo {Labb{\'e}}, Sirio {Belli}, Rychard {Bouwens}, Richard~S.
  {Ellis}, Guido {Roberts-Borsani}, Daniel~P. {Stark}, Pascal~A. {Oesch}, and
  Renske {Smit}.
\newblock {Lyman{\ensuremath{\alpha}} Emission from a Luminous z = 8.68 Galaxy:
  Implications for Galaxies as Tracers of Cosmic Reionization}.
\newblock \emph{ApJL}, 810\penalty0 (1):\penalty0 L12, September 2015.
\newblock \doi{10.1088/2041-8205/810/1/L12}.

\bibitem[{Naidu} et~al.(2022){Naidu}, {Oesch}, {van Dokkum}, {Nelson}, {Suess},
  {Whitaker}, {Allen}, {Bezanson}, {Bouwens}, {Brammer}, {Conroy},
  {Illingworth}, {Labbe}, {Leja}, {Leonova}, {Matthee}, {Price}, {Setton},
  {Strait}, {Stefanon}, {Tacchella}, {Toft}, {Weaver}, and
  {Weibel}]{Naidu_2022}
Rohan~P. {Naidu}, Pascal~A. {Oesch}, Pieter {van Dokkum}, Erica~J. {Nelson},
  Katherine~A. {Suess}, Katherine~E. {Whitaker}, Natalie {Allen}, Rachel
  {Bezanson}, Rychard {Bouwens}, Gabriel {Brammer}, Charlie {Conroy}, Garth
  {Illingworth}, Ivo {Labbe}, Joel {Leja}, Ecaterina {Leonova}, Jorryt
  {Matthee}, Sedona~H. {Price}, David~J. {Setton}, Victoria {Strait}, Mauro
  {Stefanon}, Sandro {Tacchella}, Sune {Toft}, John~R. {Weaver}, and Andrea
  {Weibel}.
\newblock {Two Remarkably Luminous Galaxy Candidates at $z\approx11-13$
  Revealed by JWST}.
\newblock \emph{arXiv e-prints}, art. arXiv:2207.09434, July 2022.

\bibitem[{Treu} et~al.(2015){Treu}, {Schmidt}, {Brammer}, {Vulcani}, {Wang},
  {Brada{\v{c}}}, {Dijkstra}, {Dressler}, {Fontana}, {Gavazzi}, {Henry},
  {Hoag}, {Huang}, {Jones}, {Kelly}, {Malkan}, {Mason}, {Pentericci},
  {Poggianti}, {Stiavelli}, {Trenti}, and {von der Linden}]{GLASS}
T.~{Treu}, K.~B. {Schmidt}, G.~B. {Brammer}, B.~{Vulcani}, X.~{Wang},
  M.~{Brada{\v{c}}}, M.~{Dijkstra}, A.~{Dressler}, A.~{Fontana}, R.~{Gavazzi},
  A.~L. {Henry}, A.~{Hoag}, K.~H. {Huang}, T.~A. {Jones}, P.~L. {Kelly}, M.~A.
  {Malkan}, C.~{Mason}, L.~{Pentericci}, B.~{Poggianti}, M.~{Stiavelli},
  M.~{Trenti}, and A.~{von der Linden}.
\newblock {The Grism Lens-Amplified Survey from Space (GLASS). I. Survey
  Overview and First Data Release}.
\newblock \emph{ApJ}, 812\penalty0 (2):\penalty0 114, October 2015.
\newblock \doi{10.1088/0004-637X/812/2/114}.

\bibitem[{Dalal} and {Kochanek}(2002)]{Dalal_2002}
N.~{Dalal} and C.~S. {Kochanek}.
\newblock {Direct Detection of Cold Dark Matter Substructure}.
\newblock \emph{ApJ}, 572\penalty0 (1):\penalty0 25--33, June 2002.
\newblock \doi{10.1086/340303}.

\bibitem[{Vegetti} et~al.(2014){Vegetti}, {Koopmans}, {Auger}, {Treu}, and
  {Bolton}]{Vegetti_2014}
S.~{Vegetti}, L.~V.~E. {Koopmans}, M.~W. {Auger}, T.~{Treu}, and A.~S.
  {Bolton}.
\newblock {Inference of the cold dark matter substructure mass function at z =
  0.2 using strong gravitational lenses}.
\newblock \emph{MNRAS}, 442\penalty0 (3):\penalty0 2017--2035, August 2014.
\newblock \doi{10.1093/mnras/stu943}.

\bibitem[{Gilman} et~al.(2020){Gilman}, {Birrer}, {Nierenberg}, {Treu}, {Du},
  and {Benson}]{Gilman_2020}
Daniel {Gilman}, Simon {Birrer}, Anna {Nierenberg}, Tommaso {Treu}, Xiaolong
  {Du}, and Andrew {Benson}.
\newblock {Warm dark matter chills out: constraints on the halo mass function
  and the free-streaming length of dark matter with eight quadruple-image
  strong gravitational lenses}.
\newblock \emph{MNRAS}, 491\penalty0 (4):\penalty0 6077--6101, February 2020.
\newblock \doi{10.1093/mnras/stz3480}.

\bibitem[{Hsueh} et~al.(2020){Hsueh}, {Enzi}, {Vegetti}, {Auger}, {Fassnacht},
  {Despali}, {Koopmans}, and {McKean}]{Hsueh_2020}
J.~W. {Hsueh}, W.~{Enzi}, S.~{Vegetti}, M.~W. {Auger}, C.~D. {Fassnacht},
  G.~{Despali}, L.~V.~E. {Koopmans}, and J.~P. {McKean}.
\newblock {SHARP - VII. New constraints on the dark matter free-streaming
  properties and substructure abundance from gravitationally lensed quasars}.
\newblock \emph{MNRAS}, 492\penalty0 (2):\penalty0 3047--3059, February 2020.
\newblock \doi{10.1093/mnras/stz3177}.

\bibitem[{Meneghetti} et~al.(2020){Meneghetti}, {Davoli}, {Bergamini},
  {Rosati}, {Natarajan}, {Giocoli}, {Caminha}, {Metcalf}, {Rasia}, {Borgani},
  {Calura}, {Grillo}, {Mercurio}, and {Vanzella}]{Meneghetti_2020}
Massimo {Meneghetti}, Guido {Davoli}, Pietro {Bergamini}, Piero {Rosati},
  Priyamvada {Natarajan}, Carlo {Giocoli}, Gabriel~B. {Caminha}, R.~Benton
  {Metcalf}, Elena {Rasia}, Stefano {Borgani}, Francesco {Calura}, Claudio
  {Grillo}, Amata {Mercurio}, and Eros {Vanzella}.
\newblock {An excess of small-scale gravitational lenses observed in galaxy
  clusters}.
\newblock \emph{Science}, 369\penalty0 (6509):\penalty0 1347--1351, September
  2020.
\newblock \doi{10.1126/science.aax5164}.

\bibitem[{Nightingale} et~al.(2022){Nightingale}, {He}, {Cao}, {Amvrosiadis},
  {Etherington}, {Frenk}, {Hayes}, {Robertson}, {Cole}, {Lange}, {Li}, and
  {Massey}]{Nightingale_2022}
James~W. {Nightingale}, Qiuhan {He}, Xiaoyue {Cao}, Aristeidis {Amvrosiadis},
  Amy {Etherington}, Carlos~S. {Frenk}, Richard~G. {Hayes}, Andrew {Robertson},
  Shaun {Cole}, Samuel {Lange}, Ran {Li}, and Richard {Massey}.
\newblock {Scanning For Dark Matter Subhalos in Hubble Space Telescope Imaging
  of 54 Strong Lenses}.
\newblock \emph{arXiv e-prints}, art. arXiv:2209.10566, September 2022.

\bibitem[Vegetti et~al.(2010)Vegetti, Koopmans, Bolton, Treu, and
  Gavazzi]{Vegetti_2010b}
S.~Vegetti, L.~V.~E. Koopmans, A.~Bolton, T.~Treu, and R.~Gavazzi.
\newblock Detection of a dark substructure through gravitational imaging.
\newblock \emph{Monthly Notices of the Royal Astronomical Society},
  408\penalty0 (4):\penalty0 1969–1981, Oct 2010.
\newblock ISSN 0035-8711.
\newblock \doi{10.1111/j.1365-2966.2010.16865.x}.
\newblock URL \url{http://dx.doi.org/10.1111/j.1365-2966.2010.16865.x}.

\bibitem[Vegetti et~al.(2012)Vegetti, Lagattuta, McKean, Auger, Fassnacht, and
  Koopmans]{Vegetti_2012}
S.~Vegetti, D.~J. Lagattuta, J.~P. McKean, M.~W. Auger, C.~D. Fassnacht, and
  L.~V.~E. Koopmans.
\newblock Gravitational detection of a low-mass dark satellite galaxy at
  cosmological distance.
\newblock \emph{Nature}, 481\penalty0 (7381):\penalty0 341–343, Jan 2012.
\newblock ISSN 1476-4687.
\newblock \doi{10.1038/nature10669}.
\newblock URL \url{http://dx.doi.org/10.1038/nature10669}.

\bibitem[{Hezaveh} et~al.(2016){Hezaveh}, {Dalal}, {Marrone}, {Mao},
  {Morningstar}, {Wen}, {Bland ford}, {Carlstrom}, {Fassnacht}, {Holder},
  {Kemball}, {Marshall}, {Murray}, {Perreault Levasseur}, {Vieira}, and
  {Wechsler}]{Hezaveh_2016}
Yashar~D. {Hezaveh}, Neal {Dalal}, Daniel~P. {Marrone}, Yao-Yuan {Mao}, Warren
  {Morningstar}, Di~{Wen}, Roger~D. {Bland ford}, John~E. {Carlstrom},
  Christopher~D. {Fassnacht}, Gilbert~P. {Holder}, Athol {Kemball}, Philip~J.
  {Marshall}, Norman {Murray}, Laurence {Perreault Levasseur}, Joaquin~D.
  {Vieira}, and Risa~H. {Wechsler}.
\newblock {Detection of Lensing Substructure Using ALMA Observations of the
  Dusty Galaxy SDP.81}.
\newblock \emph{ApJ}, 823\penalty0 (1):\penalty0 37, May 2016.
\newblock \doi{10.3847/0004-637X/823/1/37}.

\bibitem[Suyu et~al.(2020)Suyu, Huber, Ca{\~n}ameras, Kromer, Schuldt,
  Taubenberger, Y{\i}ld{\i}r{\i}m, Bonvin, Chan, Courbin,
  et~al.]{suyu2020holismokes}
SH~Suyu, S~Huber, R~Ca{\~n}ameras, M~Kromer, S~Schuldt, S~Taubenberger,
  A~Y{\i}ld{\i}r{\i}m, V~Bonvin, JHH Chan, F~Courbin, et~al.
\newblock Holismokes-i. highly optimised lensing investigations of supernovae,
  microlensing objects, and kinematics of ellipticals and spirals.
\newblock \emph{Astronomy \& Astrophysics}, 644:\penalty0 A162, 2020.

\bibitem[Birrer et~al.(2020)Birrer, Shajib, Galan, Millon, Treu, Agnello,
  Auger, Chen, Christensen, Collett, et~al.]{birrer2020tdcosmo}
S~Birrer, AJ~Shajib, A~Galan, M~Millon, T~Treu, A~Agnello, M~Auger, GC-F Chen,
  L~Christensen, T~Collett, et~al.
\newblock Tdcosmo-iv. hierarchical time-delay cosmography--joint inference of
  the hubble constant and galaxy density profiles.
\newblock \emph{Astronomy \& Astrophysics}, 643:\penalty0 A165, 2020.

\bibitem[Sérsic(1963)]{Sersic_1963}
J.~L. Sérsic.
\newblock Influence of the atmospheric and instrumental dispersion on the
  brightness distribution in a galaxy.
\newblock \emph{Boletin de la Asociacion Argentina de Astronomia La Plata
  Argentina}, 6:\penalty0 41–43, Feb 1963.

\bibitem[Brownstein et~al.(2011)Brownstein, Bolton, Schlegel, Eisenstein,
  Kochanek, Connolly, Maraston, Pandey, Seitz, Wake,
  et~al.]{brownstein2011boss}
Joel~R Brownstein, Adam~S Bolton, David~J Schlegel, Daniel~J Eisenstein,
  Christopher~S Kochanek, Natalia Connolly, Claudia Maraston, Parul Pandey,
  Stella Seitz, David~A Wake, et~al.
\newblock The boss emission-line lens survey (bells). i. a large
  spectroscopically selected sample of lens galaxies at redshift~ 0.5.
\newblock \emph{The Astrophysical Journal}, 744\penalty0 (1):\penalty0 41,
  2011.

\bibitem[Warren and Dye(2003)]{Warren_Dye_2003}
S.~J. Warren and S.~Dye.
\newblock Semilinear gravitational lens inversion.
\newblock \emph{The Astrophysical Journal}, 590:\penalty0 673–682, Jun 2003.
\newblock ISSN 0004-637X.
\newblock \doi{10/bws44f}.
\newblock URL \url{https://ui.adsabs.harvard.edu/abs/2003ApJ...590..673W}.
\newblock ADS Bibcode: 2003ApJ...590..673W.

\bibitem[Suyu et~al.(2006)Suyu, Marshall, Hobson, and Blandford]{Suyu_2006}
S.~H. Suyu, P.~J. Marshall, M.~P. Hobson, and R.~D. Blandford.
\newblock A bayesian analysis of regularized source inversions in gravitational
  lensing.
\newblock \emph{Monthly Notices of the Royal Astronomical Society},
  371:\penalty0 983–998, Sep 2006.
\newblock ISSN 0035-8711.
\newblock \doi{10/cgt4n6}.
\newblock URL \url{https://ui.adsabs.harvard.edu/abs/2006MNRAS.371..983S}.
\newblock ADS Bibcode: 2006MNRAS.371..983S.

\bibitem[Vegetti and Koopmans(2009)]{Vegetti_Koopmans_2009}
S.~Vegetti and L.~V.~E. Koopmans.
\newblock Bayesian strong gravitational-lens modelling on adaptive grids:
  objective detection of mass substructure in galaxies.
\newblock \emph{Monthly Notices of the Royal Astronomical Society},
  392\penalty0 (3):\penalty0 945–963, Jan 2009.
\newblock \doi{10.1111/j.1365-2966.2008.14005.x}.

\bibitem[Nightingale and Dye(2015)]{Nightingale_Dye_2015}
J.~W. Nightingale and S.~Dye.
\newblock Adaptive semi-linear inversion of strong gravitational lens imaging.
\newblock \emph{Monthly Notices of the Royal Astronomical Society},
  452:\penalty0 2940–2959, Sep 2015.
\newblock ISSN 0035-8711.
\newblock \doi{10/f7q7xh}.
\newblock URL \url{https://ui.adsabs.harvard.edu/abs/2015MNRAS.452.2940N}.
\newblock ADS Bibcode: 2015MNRAS.452.2940N.

\bibitem[Nightingale et~al.(2018)Nightingale, Dye, and
  Massey]{Nightingale_Dye_Massey_2018}
J.~W. Nightingale, S.~Dye, and Richard~J. Massey.
\newblock Autolens: automated modeling of a strong lens’s light, mass, and
  source.
\newblock \emph{Monthly Notices of the Royal Astronomical Society},
  478:\penalty0 4738–4784, Aug 2018.
\newblock ISSN 0035-8711.
\newblock \doi{10/gd4sks}.
\newblock URL \url{https://ui.adsabs.harvard.edu/abs/2018MNRAS.478.4738N}.
\newblock ADS Bibcode: 2018MNRAS.478.4738N.

\bibitem[Karchev et~al.(2022)Karchev, Coogan, and
  Weniger]{Karchev_Coogan_Weniger_2022}
Konstantin Karchev, Adam Coogan, and Christoph Weniger.
\newblock Strong-lensing source reconstruction with variationally optimized
  gaussian processes.
\newblock \emph{Monthly Notices of the Royal Astronomical Society},
  512:\penalty0 661–685, May 2022.
\newblock ISSN 0035-8711.
\newblock \doi{10.1093/mnras/stac311}.
\newblock URL \url{https://ui.adsabs.harvard.edu/abs/2022MNRAS.512..661K}.
\newblock ADS Bibcode: 2022MNRAS.512..661K.

\bibitem[{Mishra-Sharma} and {Yang}(2022)]{Mishra-Sharma_Yang_2022}
Siddharth {Mishra-Sharma} and Ge~{Yang}.
\newblock {Strong Lensing Source Reconstruction Using Continuous Neural
  Fields}.
\newblock \emph{arXiv e-prints}, art. arXiv:2206.14820, June 2022.

\bibitem[{Galan} et~al.(2021){Galan}, {Peel}, {Joseph}, {Courbin}, and
  {Starck}]{Galan_2021}
A.~{Galan}, A.~{Peel}, R.~{Joseph}, F.~{Courbin}, and J.~L. {Starck}.
\newblock {SLITRONOMY: Towards a fully wavelet-based strong lensing inversion
  technique}.
\newblock \emph{A\&A}, 647:\penalty0 A176, March 2021.
\newblock \doi{10.1051/0004-6361/202039363}.

\bibitem[{Birrer} et~al.(2015){Birrer}, {Amara}, and
  {Refregier}]{Birrer2015-xb}
Simon {Birrer}, Adam {Amara}, and Alexandre {Refregier}.
\newblock {Gravitational Lens Modeling with Basis Sets}.
\newblock \emph{ApJ}, 813\penalty0 (2):\penalty0 102, November 2015.
\newblock \doi{10.1088/0004-637X/813/2/102}.

\bibitem[{Birrer} and {Amara}(2018)]{Birrer2018-rq}
Simon {Birrer} and Adam {Amara}.
\newblock {lenstronomy: Multi-purpose gravitational lens modelling software
  package}.
\newblock \emph{Physics of the Dark Universe}, 22:\penalty0 189--201, December
  2018.
\newblock \doi{10.1016/j.dark.2018.11.002}.

\bibitem[{Morningstar} et~al.(2018){Morningstar}, {Hezaveh}, {Perreault
  Levasseur}, {Blandford}, {Marshall}, {Putzky}, and
  {Wechsler}]{Morningstar_2018}
Warren~R. {Morningstar}, Yashar~D. {Hezaveh}, Laurence {Perreault Levasseur},
  Roger~D. {Blandford}, Philip~J. {Marshall}, Patrick {Putzky}, and Risa~H.
  {Wechsler}.
\newblock {Analyzing interferometric observations of strong gravitational
  lenses with recurrent and convolutional neural networks}.
\newblock \emph{arXiv e-prints}, art. arXiv:1808.00011, July 2018.

\bibitem[{Morningstar} et~al.(2019){Morningstar}, {Perreault Levasseur},
  {Hezaveh}, {Blandford}, {Marshall}, {Putzky}, {Rueter}, {Wechsler}, and
  {Welling}]{Morningstar_2019}
Warren~R. {Morningstar}, Laurence {Perreault Levasseur}, Yashar~D. {Hezaveh},
  Roger {Blandford}, Phil {Marshall}, Patrick {Putzky}, Thomas~D. {Rueter},
  Risa {Wechsler}, and Max {Welling}.
\newblock {Data-driven Reconstruction of Gravitationally Lensed Galaxies Using
  Recurrent Inference Machines}.
\newblock \emph{ApJ}, 883\penalty0 (1):\penalty0 14, September 2019.
\newblock \doi{10.3847/1538-4357/ab35d7}.

\bibitem[{Adam} et~al.(2022){Adam}, {Perreault-Levasseur}, and
  {Hezaveh}]{Adam22RIM}
Alexandre {Adam}, Laurence {Perreault-Levasseur}, and Yashar {Hezaveh}.
\newblock {Pixelated Reconstruction of Gravitational Lenses using Recurrent
  Inference Machines}.
\newblock \emph{arXiv e-prints}, art. arXiv:2207.01073, July 2022.

\bibitem[{Chianese} et~al.(2020){Chianese}, {Coogan}, {Hofma}, {Otten}, and
  {Weniger}]{Chianese_2020}
Marco {Chianese}, Adam {Coogan}, Paul {Hofma}, Sydney {Otten}, and Christoph
  {Weniger}.
\newblock {Differentiable strong lensing: uniting gravity and neural nets
  through differentiable probabilistic programming}.
\newblock \emph{MNRAS}, 496\penalty0 (1):\penalty0 381--393, May 2020.
\newblock \doi{10.1093/mnras/staa1477}.

\bibitem[{Bolton} et~al.(2006){Bolton}, {Burles}, {Koopmans}, {Treu}, and
  {Moustakas}]{SLACS-1}
Adam~S. {Bolton}, Scott {Burles}, L{\'e}on V.~E. {Koopmans}, Tommaso {Treu},
  and Leonidas~A. {Moustakas}.
\newblock {The Sloan Lens ACS Survey. I. A Large Spectroscopically Selected
  Sample of Massive Early-Type Lens Galaxies}.
\newblock \emph{ApJ}, 638\penalty0 (2):\penalty0 703--724, February 2006.
\newblock \doi{10.1086/498884}.

\bibitem[{Bolton} et~al.(2008){Bolton}, {Burles}, {Koopmans}, {Treu},
  {Gavazzi}, {Moustakas}, {Wayth}, and {Schlegel}]{SLACS-5}
Adam~S. {Bolton}, Scott {Burles}, L{\'e}on V.~E. {Koopmans}, Tommaso {Treu},
  Rapha{\"e}l {Gavazzi}, Leonidas~A. {Moustakas}, Randall {Wayth}, and David~J.
  {Schlegel}.
\newblock {The Sloan Lens ACS Survey. V. The Full ACS Strong-Lens Sample}.
\newblock \emph{ApJ}, 682\penalty0 (2):\penalty0 964--984, August 2008.
\newblock \doi{10.1086/589327}.

\bibitem[{Shu} et~al.(2017){Shu}, {Brownstein}, {Bolton}, {Koopmans}, {Treu},
  {Montero-Dorta}, {Auger}, {Czoske}, {Gavazzi}, {Marshall}, and
  {Moustakas}]{SLACS-13}
Yiping {Shu}, Joel~R. {Brownstein}, Adam~S. {Bolton}, L{\'e}on V.~E.
  {Koopmans}, Tommaso {Treu}, Antonio~D. {Montero-Dorta}, Matthew~W. {Auger},
  Oliver {Czoske}, Rapha{\"e}l {Gavazzi}, Philip~J. {Marshall}, and Leonidas~A.
  {Moustakas}.
\newblock {The Sloan Lens ACS Survey. XIII. Discovery of 40 New Galaxy-scale
  Strong Lenses}.
\newblock \emph{ApJ}, 851\penalty0 (1):\penalty0 48, December 2017.
\newblock \doi{10.3847/1538-4357/aa9794}.

\bibitem[{Brownstein} et~al.(2012){Brownstein}, {Bolton}, {Schlegel},
  {Eisenstein}, {Kochanek}, {Connolly}, {Maraston}, {Pandey}, {Seitz}, {Wake},
  {Wood-Vasey}, {Brinkmann}, {Schneider}, and {Weaver}]{BELLS-1}
Joel~R. {Brownstein}, Adam~S. {Bolton}, David~J. {Schlegel}, Daniel~J.
  {Eisenstein}, Christopher~S. {Kochanek}, Natalia {Connolly}, Claudia
  {Maraston}, Parul {Pandey}, Stella {Seitz}, David~A. {Wake}, W.~Michael
  {Wood-Vasey}, Jon {Brinkmann}, Donald~P. {Schneider}, and Benjamin~A.
  {Weaver}.
\newblock {The BOSS Emission-Line Lens Survey (BELLS). I. A Large
  Spectroscopically Selected Sample of Lens Galaxies at Redshift
  \raisebox{-0.5ex}\textasciitilde0.5}.
\newblock \emph{ApJ}, 744\penalty0 (1):\penalty0 41, January 2012.
\newblock \doi{10.1088/0004-637X/744/1/41}.

\bibitem[{Shu} et~al.(2016){Shu}, {Bolton}, {Kochanek}, {Oguri},
  {P{\'e}rez-Fournon}, {Zheng}, {Mao}, {Montero-Dorta}, {Brownstein},
  {Marques-Chaves}, and {M{\'e}nard}]{BELLS-3}
Yiping {Shu}, Adam~S. {Bolton}, Christopher~S. {Kochanek}, Masamune {Oguri},
  Ismael {P{\'e}rez-Fournon}, Zheng {Zheng}, Shude {Mao}, Antonio~D.
  {Montero-Dorta}, Joel~R. {Brownstein}, Rui {Marques-Chaves}, and Brice
  {M{\'e}nard}.
\newblock {The BOSS Emission-line Lens Survey. III. Strong Lensing of
  Ly{\ensuremath{\alpha}} Emitters by Individual Galaxies}.
\newblock \emph{ApJ}, 824\penalty0 (2):\penalty0 86, June 2016.
\newblock \doi{10.3847/0004-637X/824/2/86}.

\bibitem[{Collett}(2015)]{Collett_2015}
Thomas~E. {Collett}.
\newblock {The Population of Galaxy-Galaxy Strong Lenses in Forthcoming Optical
  Imaging Surveys}.
\newblock \emph{ApJ}, 811\penalty0 (1):\penalty0 20, September 2015.
\newblock \doi{10.1088/0004-637X/811/1/20}.

\bibitem[{Ramesh} et~al.(2022){Ramesh}, {Dhariwal}, {Nichol}, {Chu}, and
  {Chen}]{dalle2}
Aditya {Ramesh}, Prafulla {Dhariwal}, Alex {Nichol}, Casey {Chu}, and Mark
  {Chen}.
\newblock {Hierarchical Text-Conditional Image Generation with CLIP Latents}.
\newblock \emph{arXiv e-prints}, art. arXiv:2204.06125, April 2022.

\bibitem[{Bond-Taylor} et~al.(2021){Bond-Taylor}, {Leach}, {Long}, and
  {Willcocks}]{generative-review}
Sam {Bond-Taylor}, Adam {Leach}, Yang {Long}, and Chris~G. {Willcocks}.
\newblock {Deep Generative Modelling: A Comparative Review of VAEs, GANs,
  Normalizing Flows, Energy-Based and Autoregressive Models}.
\newblock \emph{arXiv e-prints}, art. arXiv:2103.04922, March 2021.

\bibitem[Sohl-Dickstein et~al.(2015)Sohl-Dickstein, Weiss, Maheswaranathan, and
  Ganguli]{Sohl-Dickstein_2015}
Jascha Sohl-Dickstein, Eric~A. Weiss, Niru Maheswaranathan, and Surya Ganguli.
\newblock Deep unsupervised learning using nonequilibrium thermodynamics, 2015.
\newblock URL \url{https://arxiv.org/abs/1503.03585}.

\bibitem[Ho et~al.(2020)Ho, Jain, and Abbeel]{ddpm}
Jonathan Ho, Ajay Jain, and Pieter Abbeel.
\newblock Denoising diffusion probabilistic models, 2020.
\newblock URL \url{https://arxiv.org/abs/2006.11239}.

\bibitem[Smith et~al.(2022)Smith, Geach, Jackson, Arora, Stone, and
  Courteau]{astroddpm}
Michael~J Smith, James~E Geach, Ryan~A Jackson, Nikhil Arora, Connor Stone, and
  St{\'{e}}phane Courteau.
\newblock Realistic galaxy image simulation via score-based generative models.
\newblock \emph{Monthly Notices of the Royal Astronomical Society},
  511\penalty0 (2):\penalty0 1808--1818, jan 2022.
\newblock \doi{10.1093/mnras/stac130}.
\newblock URL \url{https://doi.org/10.1093%2Fmnras%2Fstac130}.

\bibitem[Kawar et~al.(2022)Kawar, Elad, Ermon, and Song]{ddrm}
Bahjat Kawar, Michael Elad, Stefano Ermon, and Jiaming Song.
\newblock Denoising diffusion restoration models, 2022.
\newblock URL \url{https://arxiv.org/abs/2201.11793}.

\bibitem[Meneghetti(2021)]{Meneghetti-lensing}
Massimo Meneghetti.
\newblock \emph{Introduction to Gravitational Lensing}, volume 956 of
  \emph{Lecture Notes in Physics}.
\newblock Springer Nature, 2021.
\newblock ISBN 978-3-030-73581-4.
\newblock URL \url{https://link.springer.com/book/10.1007/978-3-030-73582-1}.

\bibitem[Kawar et~al.(2021)Kawar, Vaksman, and Elad]{snips}
Bahjat Kawar, Gregory Vaksman, and Michael Elad.
\newblock Snips: Solving noisy inverse problems stochastically.
\newblock In \emph{Advances in Neural Information Processing Systems}, Oct
  2021.
\newblock URL \url{https://openreview.net/forum?id=pBKOx_dxYAN}.

\bibitem[Stone and Courteau(2019)]{Stone_2019}
Connor Stone and St{\'{e} }phane Courteau.
\newblock The intrinsic scatter of the radial acceleration relation.
\newblock \emph{The Astrophysical Journal}, 882\penalty0 (1):\penalty0 6, aug
  2019.
\newblock \doi{10.3847/1538-4357/ab3126}.
\newblock URL \url{https://doi.org/10.3847%2F1538-4357%2Fab3126}.

\bibitem[Stone et~al.(2021)Stone, Courteau, and Arora]{Stone_2021}
Connor Stone, St{\'{e} }phane Courteau, and Nikhil Arora.
\newblock The intrinsic scatter of galaxy scaling relations.
\newblock \emph{The Astrophysical Journal}, 912\penalty0 (1):\penalty0 41, may
  2021.
\newblock \doi{10.3847/1538-4357/abebe4}.
\newblock URL \url{https://doi.org/10.3847%2F1538-4357%2Fabebe4}.

\bibitem[Friedman(2003)]{Friedman:2003id}
Jerome~H. Friedman.
\newblock {On multivariate goodness of fit and two sample testing}.
\newblock \emph{eConf}, C030908:\penalty0 THPD002, 2003.

\bibitem[Lopez-Paz and Oquab(2017)]{lopez2017revisiting}
David Lopez-Paz and Maxime Oquab.
\newblock Revisiting classifier two-sample tests.
\newblock In \emph{International Conference on Learning Representations}, 2017.

\end{thebibliography}
\endgroup

%%%%%%%%%%%%%%%%%%%%%%%%%%%%%%%%%%%%%%%%%%%%%%%%%

\begin{appendix}

\section{Appendix: \texorpdfstring{\acrfull*{ddrm}}{DDRM}}\label{apx:ddrm}

Starting with the \gls*{svd} of $\H = \U \S \Vh$ and the transformed observation $\bar{\y} \equiv \S^{+}\U^{\transpose}\y$ (with $^{+}$ a Moore-Penrose pseudo inverse), \gls*{ddrm} consists of applying the following updates:
\begin{alignat}{2}
    & \denoisep[T](\xbari_T \given \y) && = \begin{cases}
        \opbraces{\Normal}(0, \sigma_T^2)
        & \text{if $\si = 0$},
        \\
        \opbraces{\Normal}(\ybari, \sigma_T^2 - \frac{\noise^2}{\si^2})
        & \text{if $\si > 0$};
    \end{cases}
    \label{eqn:ddrm-1}
    \\
    \label{eqn:ddrm-2}
    & \denoisep(\xbari_t \given \x_{t+1}, \y) && = \begin{cases}
        \opbraces{\Normal}(
            \xrec + \sqrt{1-\eta^2} \frac{\sigma_t}{\sigma_{t+1}} \qty(\xbari_{t+1} - \xrec),
            \eta^2 \sigma_t^2
        ) & \text{if $\si = 0$},
        \\
        \opbraces{\Normal}(
            \xrec + \sqrt{1-\eta^2} \frac{\sigma_t}{\noise / \si} \qty(\ybari - \xrec),
            \eta^2 \sigma_t^2
        ) & \text{if $\sigma_t < \noise / \si$},
        \\
        \opbraces{\Normal}(
            \qty(1-\eta_b) \xrec + \eta_b \ybari,
            \sigma_t^2 - \eta_b^2\frac{\noise^2}{\si^2}
        ) & \text{if $\sigma_t \geq \noise / \si$}.
    \end{cases}
\end{alignat}
Here $\mathbf{\cdot}^{(i)}$ labels the $i$\textsuperscript{th} component of a vector. At the beginning of each iteration, the current transformed prediction $\xbar_{t+1}$ is de-rotated into $\x_{t+1} = \V \xbar_{t+1}$, which is then denoised: $\x_{\nnparams, t} = \meanfunc(\x_{t+1})$, and rotated back into $\xbar_{\nnparams, t} \equiv \Vh \x_{\nnparams, t}$.

\Cref{eqn:ddrm-2} allows for controlling the relative information content carried by noise versus that encoded in the network: when $\eta = 1$, unconstrained pixels (first case) are sampled independently at each denoising step, whereas setting $\eta=0$ connects them deterministically to the initial noise realisation. Furthermore, in high-noise scenarios, which correspond to the second case of \cref{eqn:ddrm-2}, $\eta$ controls how strongly denoising is steered towards the particular observation, with low values leading to stronger conditioning.

\citet{ddrm} prove the equivalence of the \gls*{ddrm} and \gls*{ddpm} \gls*{elbo} objectives, which allows one to use a pre-trained unconditioned \gls*{ddpm} model as a denoiser in \gls*{ddrm}, under the condition $\eta=1$. They show that for other choices of $\eta$ (and even of $\eta_b$, which may also be considered a hyperparameter), the objectives remain similar, so \emph{approximate} \gls*{ddrm} can still be performed.

\begin{figure}[p]
    \centering
    \includegraphics[width=\linewidth]{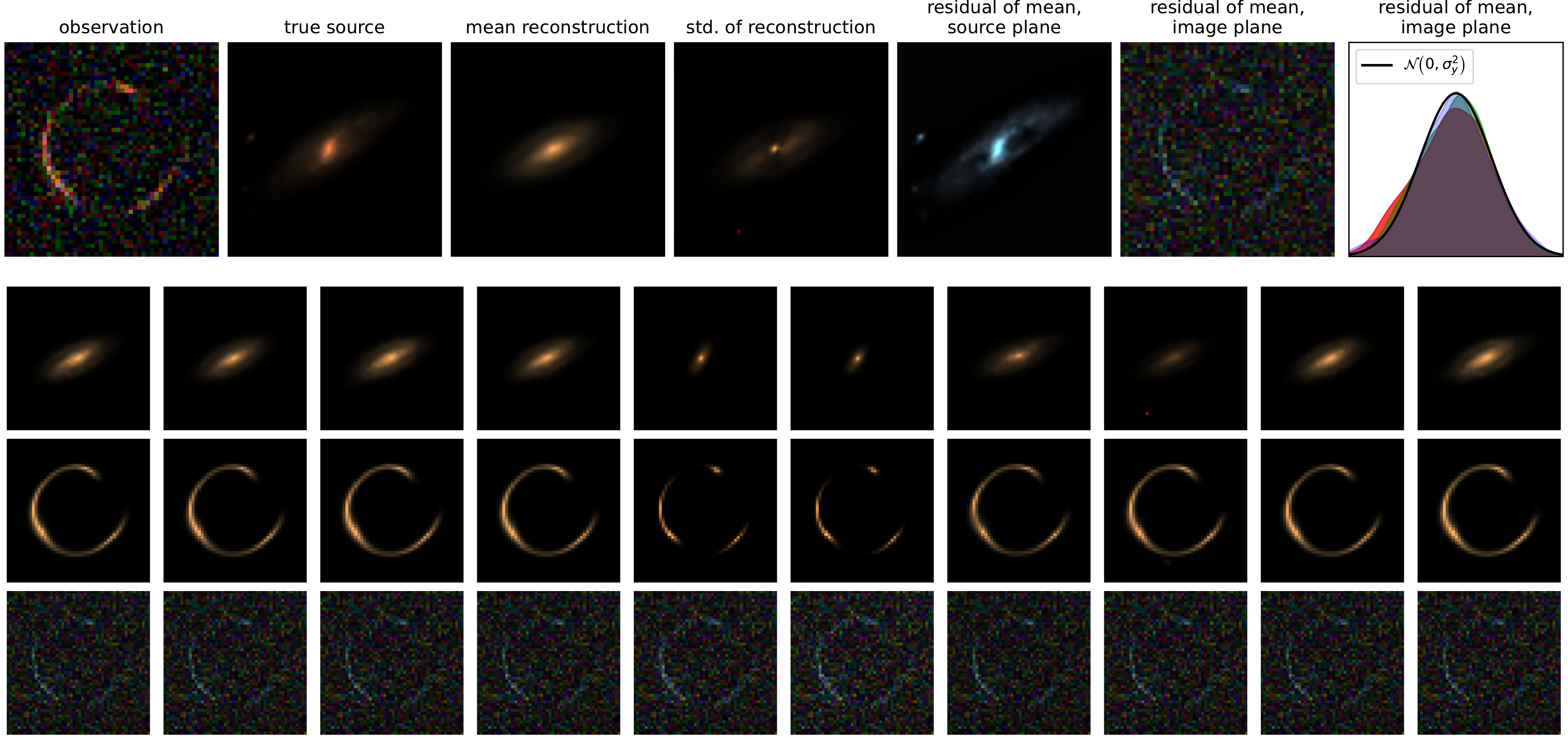}
    \caption{Same as \cref{fig:results-med}, but with the high noise setting (peak signal-to-noise ratio \num{6}).\label{fig:results-high}}
\end{figure}
\begin{figure}[p]
    \centering
    \includegraphics[width=\linewidth]{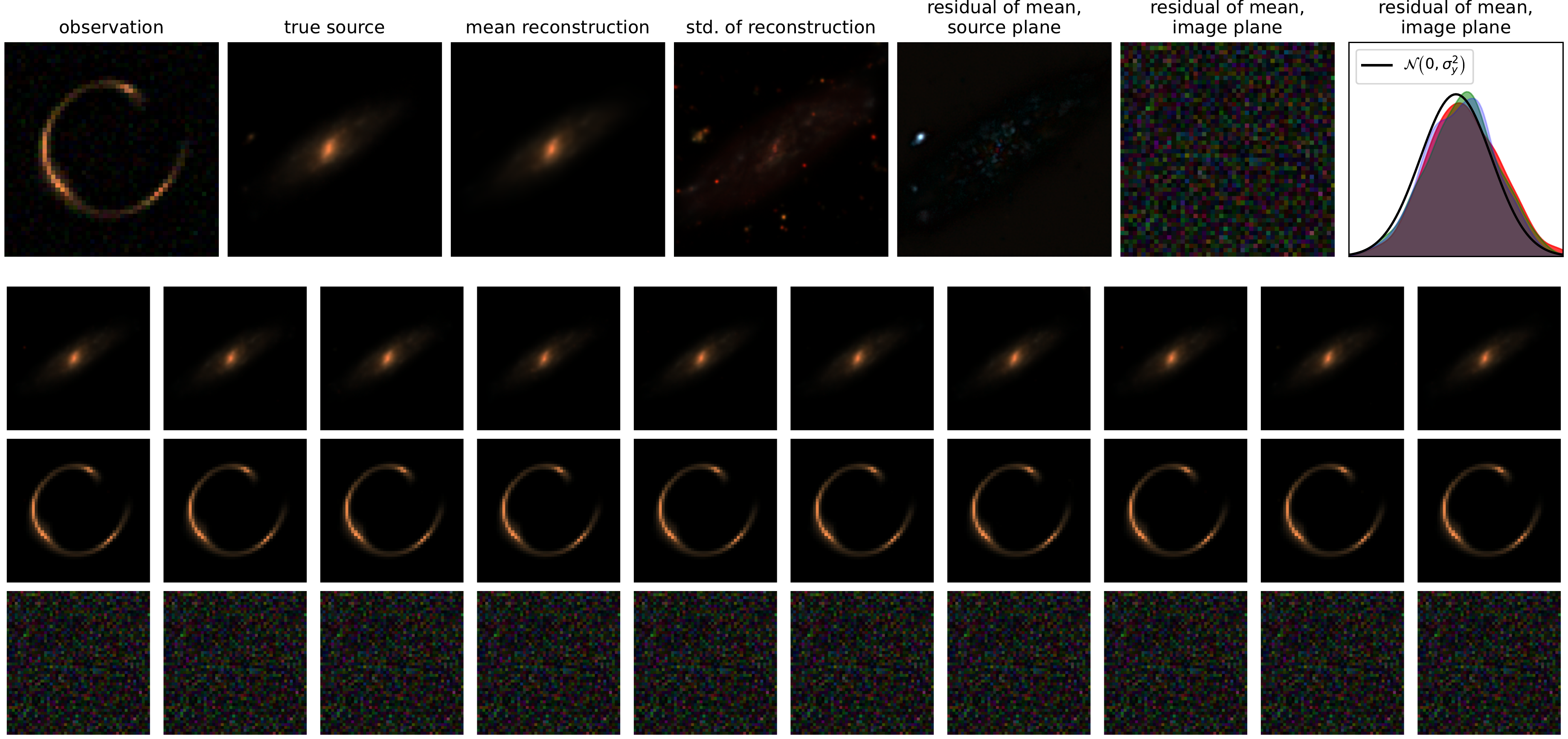}
    \caption{Same as \cref{fig:results-med} (medium-noise setting), but inference has been performed with $\eta = \num{0.03}$.\label{fig:results-hyper-med}}
\end{figure}
\begin{figure}[p]
    \centering
    \includegraphics[width=\linewidth]{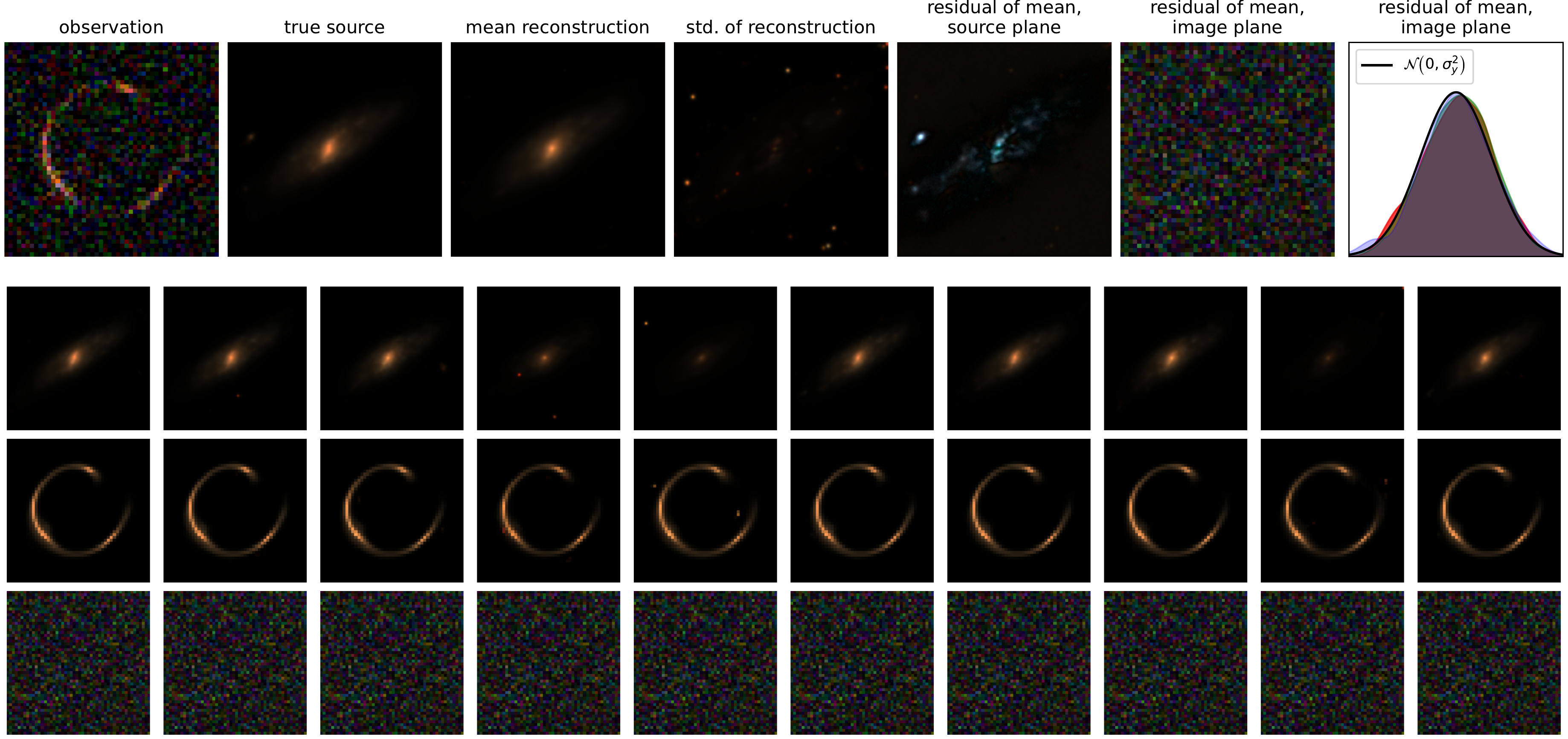}
    \caption{Same as \cref{fig:results-high} (high-noise setting), but inference has been performed with $\eta = \num{0.03}$.\label{fig:results-hyper-high}}
\end{figure}

In \cref{fig:results-hyper-med,fig:results-hyper-high} we briefly explore the effect setting a low $\eta=\num{0.03}$ has on reconstructions. In the medium-noise scenario, \cref{fig:results-hyper-med}, results are similar to using $\eta=1$, but now the generative process produces artefacts like spots, which are common in unconditioned \AstroDDPM samples (see fig.~2 of \citet{astroddpm}) but unwarranted by data. In the high-noise setting, \cref{fig:results-hyper-high}, residuals are much improved from the case of $\eta=1$ due to the stronger conditioning on the observation. These qualitative tests show the importance of tuning $\eta$ so as to match the regime in which \AstroDDPM has been trained (in fact, \citet{astroddpm} use the equivalence between \gls*{ddpm} and score-matching described in \citet{ddpm} to train their model, so they do not need to explicitly set a parameter like $\eta$). We plan to optimise $\eta$ in the future by quantitatively measuring the quality of posterior samples with the classifier 2-sample test \citep{Friedman:2003id,lopez2017revisiting}.

\end{appendix}

\end{document}